\documentclass[preprint, 11pt, authoryear]{elsarticle}

\usepackage{lipsum}
\makeatletter
\def\ps@pprintTitle{%
	\let\@oddhead\@empty
	\let\@evenhead\@empty
	\def\@oddfoot{}%
	\let\@evenfoot\@oddfoot}
\makeatother

\usepackage{setspace}
\usepackage{psfrag}
\usepackage{natbib}
\usepackage{graphicx}
\usepackage{caption}
\usepackage{subcaption}
\usepackage{hyperref}
\usepackage{enumitem}
\usepackage{color}
\usepackage{amsmath}
\usepackage{amssymb}
\usepackage{amsfonts, amsmath, amsthm, amssymb}
\usepackage{fullpage}
\usepackage{lineno}

\newtheorem{theorem}{Theorem}[section]
\newtheorem{remark}[theorem]{Remark}
\usepackage{amssymb}

\begin{document}

\begin{frontmatter}



\title{A thermodynamic approach to nonlinear ultrasonics for material state awareness and prognosis}


\author[lanl] {Vamshi Krishna Chillara \corref{cor1}}
 
\address[lanl]{Materials Physics and Applications, MPA 11, Los Alamos National Laboratory, NM 87545}
 
\ead{vchillara@lanl.gov; Ph:814-954-2291}

\cortext[cor1]{Corresponding author}

\begin{abstract}
We develop a thermodynamic framework for modeling nonlinear ultrasonic damage sensing and prognosis in materials undergoing progressive damage. The framework is based on the internal variable approach and relies on the construction of a pseudo-elastic strain energy function that captures the energetics associated with the damage progression. The pseudo-elastic strain energy function is composed of two energy functions --- one that describes how a material stores energy in an elastic fashion and the other describes how material dissipates energy or stores it in an inelastic fashion. Experimental motivation for the choice of the above two functionals is discussed and some specific choices pertaining to damage progression during fatigue and creep are presented. The thermodynamic framework is employed to model the nonlinear response of material undergoing stress relaxation and creep-like degradation. For each of the above cases, evolution of the nonlinearity parameter with damage as well as with macroscopic measurables like accumulated plastic strain are obtained.           
\end{abstract}

\end{frontmatter}

\section{Introduction}
Predictive structural health monitoring methodologies to continuously monitor the state of structures and predict their remaining useful life are widely being researched with the goal of achieving a paradigm shift from scheduled based maintenance to condition based maintenance. These are intended to ensure a reduction in life-cycle costs and enhanced structural safety. In this context, damage detection methodologies to detect and characterize damage early in its lifetime are receiving considerable attention. Of these, nonlinear ultrasonic techniques \citep{jhang2009nonlinear} are being  widely employed due to their applicability to a variety of materials ranging from metals, composites to bio-materials like tissues. Some of the early investigations concerning the use of nonlinear ultrasonics to characterize damage in metals can be traced back to the works of \citet{breazeale1963finite,breazeale1965ultrasonic, hikata1965dislocation,hikata1966generationa,hikata1966generationb, blackburn1984nonlinear}. They employed second harmonic generation from ultrasonic waves to quantify dislocation density in metals. Cantrell and co-authors \citep{cantrell2001nonlinear} employed second harmonic generation to characterize fatigue damage and also developed dislocation based theories \citep{cantrell2004substructural,cantrell2006quantitative} to quantify acoustic nonlinearity in metals. These works spurred significant interest into using nonlinear ultrasonics  to  characterize damage progression during creep \citep{baby2008creep}, radiation damage \citep{matlack2012evaluation}, etc. A comprehensive review of the theoretical and experimental aspects of nonlinear ultrasonic techniques for early damage characterization is presented in \citet{matlack2014review}. It should be emphasized that the above techniques rely on classical nonlinearity of material response. Other class of nonlinear techniques that rely on nonclassical nonlinearity \citep{delsanto2006universality} of material response include Nonlinear elastic wave spectroscopy (NEWS) \citep{van2000nonlinearb,van2000nonlineara,van2001micro}, Nonlinear resonant ultrasound spectroscopy (NRUS) \citep{muller2005nonlinear}, etc. Both classical and nonclassical nonlinear material responses have extensively been studied and several experimental investigations have been carried out. However, most of these studies were aimed at nondestructive assessment of damage and further research needs to be carried out to employ these for online monitoring of structures. In this regard, fusion of nonlinear ultrasonic methodologies with guided wave inspection schemes are receiving increasing attention \citep{chillara2016review}. These, henceforth referred to as ``nonlinear guided waves'' combine the advantages of nonlinear ultrasonics with guided wave inspection schemes for continuous monitoring of structures. Several theoretical \citep{de2003finite,deng1999cumulative,muller2010characteristics,matsuda2011phase, chillara2012interaction}, numerical \citep{chillara2014nonlinear,chillara2014qnde,leamy2014local,rauter2015numerical},  and experimental investigations \citep{bermes2007experimental,deng2007assessment,lissenden2014effect,lissenden2014nonlinear,rauter2015impact} that explore the feasibility of using nonlinear guided waves for early damage detection have been carried out. While the above nonlinear techniques appear promising for nondestructive assessment of damage, there is a need to develop a quantitative understanding  of micromechanisms responsible for damage progression.

Hikata and co-workers   \citep{hikata1965dislocation,hikata1966generationa,hikata1966generationb} were one of the first to develop models that relate dislocation density to nonlinear material behavior. These were based on string model for dislocations originally developed in \citet{eshelby1949dislocations} and \citet{granato1956theory}. Along the same lines, Cantrell \citep{cantrell2004substructural} quantified the contribution of dislocation monopoles and dipoles to acoustic nonlinearity. These models were widely used to quantify nonlinear response of metallic materials \citep{xiang2012effect,xiang2014creep} undergoing progressive damage. Despite the widespread use of dislocation-based models, it should be recognized that the ultrasonic waves are only sensitive to the overall effect of micro-scale damage over a length scale corresponding to their wavelength. Hence, the above models that rely on the interaction of waves with  dislocations need to be scaled-up to the meso-scale using homogenization schemes. The need for such an approach was highlighted in \citet{chillara2014ijes,chillara2015spie}. Such homogenization approaches for quantifying nonlinear behavior due to microcracks \citep{nazarov1997nonlinear, {zhao2015frequency}} and microvoids \cite{chillara2015spie} were developed. These models invariably employ some measures that quantify the internal damage; volume-fraction of defects is one such example. In other words, the resulting homogenized model is a continuum model that incorporates the internal damage state of the material. This notion of continuum models with internal variables was generalized \citep{maugin1994thermodynamics,maugin1994thermodynamics2} for developing models pertaining to a plethora of applications like plasticity, coupled processes like chemo-mechanics, etc. One widespread use of this approach is in modeling degradation of materials.      

The goal of this article is to develop a framework based on the internal variable approach  to model the nonlinear response of materials undergoing progressive damage. The framework relies on the notion of the material-state as captured by a (internal) damage variable. The energetics associated with damage growth is captured using two energy functions; one for describing the way the material stores energy in an elastic fashion and the other for describing the way the material dissipates energy or stores it in an inelastic fashion. The motivation for such a framework stems from the need to develop experimentally driven models for damage sensing and prognosis --- two important aspects of structural health monitoring. In particular, the framework enables one to utilize the existing experimental data pertaining to nondestructive evaluation and characterization of materials (usually carried out in laboratory conditions) and extend them for damage prognosis in structural health monitoring applications. The key feature of this approach is that it results in models that capture the evolution of the macroscopic damage measurables like plastic strain and material nonlinearity --- the latter one is of particular interest to this article as material nonlinearity is generally accepted to be an early indicator of damage in structures. We envisage the above framework to be applicable to a certain class of  mechanical, thermo-mechanical and chemo-mechanical degradations.     

The content of this article is organized as follows. Section 2 presents the constitutive framework for modeling degradation. We discuss specific choices for the constitutive response functions with the focus being on modeling the nonlinear ultrasonic response under various damage progression scenarios. Section 3 presents examples where the above framework is employed to model nonlinear response in materials undergoing stress-relaxation and creep-like degradation. Finally, we present conclusions in section 4.

\section{Constitutive framework for nonlinear response of materials undergoing progressive damage: an internal variable approach \label{sec2}}

In this section, we present the constitutive framework for  modeling nonlinear response of solids undergoing progressive degradation. We employ the internal variable approach for modeling the material response. We first discuss the notation in section \ref{sec2.1} followed by the constitutive framework in section \ref{sec2.2} and choice of constitutive response functions for modeling damage in section \ref{sec2.3}.  

\subsection{Notation \label{sec2.1}}
 Let ${\bf X}$ denote the position of the material particle in the reference configuration and ${\bf x}$ denote its position in the current configuration. We denote the deformation gradient by $\bf F=\frac{\partial x}{\partial X}$, the displacement gradient by $\bf H$ and the Lagrangian strain by $\bf E$. The following relations exist between these kinematic variables, where {$\bf I$} denotes the identity tensor.

\begin{equation}{\label{eqn1}}
\bf F=I+H
\end{equation}

\begin{equation}{\label{eqn2}}
{\bf E}=  \frac{1}{2} {\bf(F^TF-I)}=  \frac{1}{2} {\bf(H+H^T+H^TH)}.
\end{equation}
The second Piola-Kirchhoff stress is denoted by $\bf T_{RR}$. For hyperelastic materials with strain energy density $\textrm W$, we have
\begin{equation} {\label{eqn3}}
\bf T_{RR}=\bf \frac{\partial \textrm{W}}{\partial E}.
\end{equation}
Likewise, the first Piola-Kirchhoff stress tensor denoted by $\bf S$ is given by 
\begin{equation} {\label{eqn3a}}
\bf S=\bf \frac{\partial \textrm{W}}{\partial F}.
\end{equation}
The first and second Piola-Kirchhoff stresses are related by
\begin{equation} {\label{eqn3b}}
\bf S=\bf FT_{RR} .
\end{equation}
Next, we discuss the constitutive framework.

\subsection{Constitutive framework \label{sec2.2}}

As highlighted earlier, the framework is based on the internal variable approach and continuum damage mechanics can be considered a sub-class of this general approach. We start with a list of assumptions that form the basis for the proposed framework.

\begin{enumerate}
\item We assume that the state of the material is characterized by a set of (internal) variables $\{\Gamma_1,\Gamma_2,\cdots,\Gamma_n\}$ that is collectively denoted by $\Gamma$. For example, for a material undergoing creep, one may consider $\Gamma_1$ to correspond to the void fraction in the material. Likewise, for a material undergoing chemo-mechanical degradation, $\Gamma_2$ may represent the concentration of infusing chemical \citep{mudunuru2012framework, xu2015material}. Hence, the number and nature of variables used depend on the physical processes at hand. However, it is desired that these variables are measurables either from a macroscopic or microscopic standpoint.              

\item \textit{For each material state $\Gamma$, we associate a reference configuration for the material from which its mechanical response is assumed to be hyperelastic with a strain energy function $\mathrm W_{el}({\bf E},\Gamma)=\mathrm W_{el}({\bf E},\Gamma_1,\Gamma_2,\cdots,\Gamma_n)$ as introduced in \citet{chillara2014ijes}. Note that the assumption implies that for a given set of constants $\Gamma$, the material can only respond in an elastic fashion i.e., without energy dissipation. In other words, we restrict ourselves to elastic deformations from a given material state. This is a reasonable assumption when we are interested in the response of material to ultrasonic waves or small amplitude vibrations that fall well within the elastic regime\footnote{We are interested in understanding damage growth as characterized by ultrasound. There are two different dynamic processes involved --- one corresponding to damage growth and the other due to ultrasonic waves used to probe the material. These have disparate time-scales (See \ref{rem3})}. Neverthless, energy dissipation can occur through a change in $\Gamma$ i.e., when a material changes state due to microstructural changes as would be the case in materials undergoing creep or fatigue damage.}

\item The set $\Gamma=\{\Gamma_1,\Gamma_2,\cdots,\Gamma_n\}$ may evolve with time as in the case of a material undergoing progressive microstructural changes or degradation.

\item In this article, we restrict ourselves to materials and deformations that are homogeneous and hence $\Gamma$ is assumed to be independent of spatial variables.
\end{enumerate} 
Under the above assumptions, material state awareness involves quantifying the (internal) variables ($\Gamma$) and understanding how they evolve. In other words, quantifying $\Gamma$ entails damage sensing and understanding how $\Gamma$ evolves entails damage prognosis. Next, we discuss our framework based on the above assumptions.

We begin our discussion by defining a pseudo-elastic energy density (per unit reference volume) function $\mathrm W({\bf{E}},\Gamma)$ \citep{ogden1999pseudo} as follows
\begin{equation} \label{eqn4}
\textrm W({\bf{E}},\Gamma)=\textrm W_{el}({\bf{E}},\Gamma)+\textrm W_{nel}(\Gamma).
\end{equation} 
While $\textrm W_{el}({\bf{E}},\Gamma)$ denotes the elastic energy stored in the material and is recoverable, $\textrm W_{nel}(\Gamma)$ denotes the total non-recoverable energy that is either stored in the material or dissipated. Note that $\textrm W_{nel}(\Gamma)$ is assumed to depend only on $\Gamma$. Moreover, if we denote the virgin state of the material by $\Gamma^0$ and set $\textrm W_{nel}(\Gamma^0)=0$, it is easy to see that
\begin{equation} \label{eqn5}
\textrm W({\bf{0}},\Gamma)-\textrm W({\bf{0}},\Gamma^0)=\textrm W_{nel}(\Gamma)
\end{equation}
because $\textrm W_{el}({\bf{E}},\Gamma)$ is fully recoverable.
  
In other words, $\textrm W({\bf 0}, \Gamma)$ characterizes the amount of energy expended in taking the system from the state $\Gamma^0$ to $\Gamma$ and is assumed to be independent of the rate at which this process has occurred. A more general framework can be obtained by  considering the energy expended to depend on the rate of the process \citep{rajagopal2007shear}.  However, the definition as in Eqn(\ref{eqn4}) suffices for the purpose of this article where we assume the damage evolution to  be quasi-static ($\dot{\Gamma} \approx 0$).

\begin{remark}
	It is worth noting that the functional  $\mathrm{W}({\bf{E}},\Gamma)$ is a state function in $({\bf E},\Gamma)$ state space. Moreover, a material in state $({\bf E}, \Gamma)$ can be elastically unloaded to $({\bf 0}, \Gamma)$ i.e., without dissipation and the response during unloading is characterized by ${\mathrm W}_{el}({\bf{E}},\Gamma)$.	
\end{remark}
  
From the above remark, it follows that the equations of equilibrium for the material can be obtained using standard variational mechanics principles \citep{lazopoulos1998nonlinear} and are given by

\begin{subequations}
\begin{eqnarray} 
\label{eqn6} \textrm{Div}(\textrm {\bf S}( {\bf E},\Gamma))&=&{\bf 0}  \\
\label{eqn6_1} \frac{\partial \textrm W({\bf E},\Gamma)}{\partial \Gamma}&=&{\bf 0}
\end{eqnarray}
\end{subequations}
with appropriate boundary conditions.
\begin{remark}
	Equations (\ref{eqn6} \& \ref{eqn6_1}) are obtained under the assumption that $W({\bf E},\Gamma)$ is sufficiently smooth. 	
\end{remark}

We would like to emphasize that the construction of the pseudo-elastic energy function $\textrm W({\bf{E}},\Gamma)$ is actually motivated by experiments. In particular, most of the experiments concerning ultrasonic characterization of materials undergoing progressive degradation are carried out in the unloaded configuration. For example, in the case of fatigue, the material is cyclically loaded in a tensile testing machine for different number of cycles and the ultrasonic characterization of the specimen is carried out when it is unloaded and removed from the test machine. In doing so, we are actually changing the state of the specimen from some $({\bf E},\Gamma)$ to $({\bf 0},\Gamma)$. Once the material is in the unloaded state, ultrasonic waves characterize the response of the material governed only by ${\mathrm W}_{el}({\bf{E}},\Gamma)$. However, another significant piece of information concerning material state is in $\textrm W_{nel}(\Gamma)$ and plays a key role in damage progression as will be discussed later. 

\begin{figure}[p]
 	\centering
 	\psfrag{A}{$\Gamma^0$}
 	\psfrag{A1}{$\Gamma^1$(${\epsilon_p}^1$)}
 	\psfrag{A2}{$\Gamma^2$(${\epsilon_p}^2$)}
 	\psfrag{B}{$\sigma$}
 	\psfrag{C}{$\epsilon$}
 	\psfrag{S}{$\sigma_y$}
 	\includegraphics[scale=1]{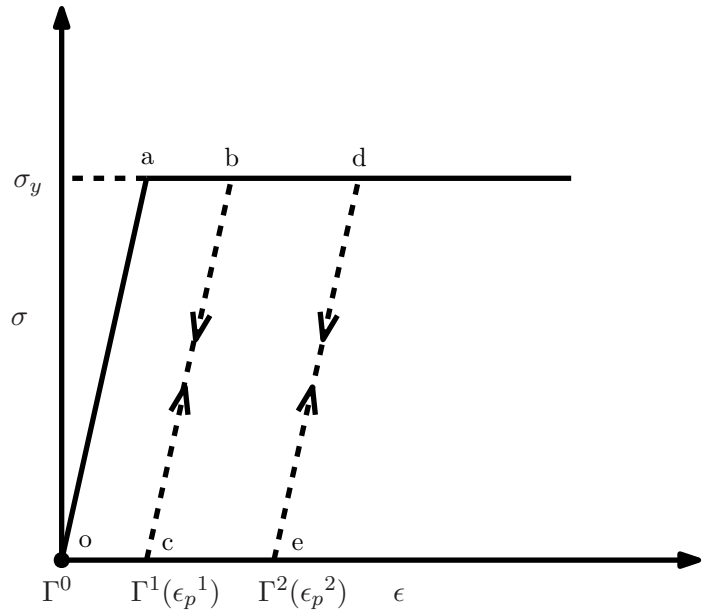}
 	\caption{Stress-strain curves for an elastic-perfectly-plastic material depicting different material states. \label{fig1}}
 \end{figure}

Figure \ref{fig1} illustrates the above introduced notions of material state ($\Gamma$), ${\mathrm W}_{el}({\bf{E}},\Gamma)$ and ${\mathrm W}_{nel}(\Gamma)$ for the case of an elastic-perfectly-plastic material. Suppose that one is performing an experiment wherein a material in the virgin state $\Gamma^0$ is plastically (uniaxially) loaded to a prescribed plastic strain and then (elastically) unloaded so that the material attains a different state $\Gamma^1$. Likewise, loading it to a different strain and unloading would take it to $\Gamma^2$. Each of the states $\Gamma^1$ and $\Gamma^2$ can be uniquely characterized by the plastic strains ${\epsilon_p}^1$ and ${\epsilon_p}^2$ in the specimen upon unloading. Moreover, ${\mathrm W}_{el}({\bf{E}},\Gamma_1)$ and ${\mathrm W}_{el}({\bf{E}},\Gamma_2)$ characterize the elastic response of the material along lines `\textit{bc}' and `\textit{de}' respectively. Note that the elastic strain ${\bf E}$ in  ${\mathrm W}_{el}({\bf{E}},\Gamma_1)$ and ${\mathrm W}_{el}({\bf{E}},\Gamma_2)$ is referred to from the unloaded configurations $\Gamma_1$ and $\Gamma_2$ respectively. If we assume the yield strength of the material to be $\sigma_y$, ${\mathrm W}_{nel}(\Gamma_1)$ is the energy dissipated in taking the material from $\Gamma_0$ to $\Gamma_1$ and is given by area under the curve `\textit{oabc}' i.e., $|\sigma_y| |{\epsilon_p}^1| $. So, the pseudo elastic strain energy function for the elastic-perfectly-plastic material is given by 

\begin{equation}\label{eqn6a}
\textrm W({\bf{E}},\Gamma)=\textrm W_{el}({\bf{E}},\Gamma)+|\sigma_y| |{\epsilon_p}|
\end{equation}   
where ${\epsilon_p}$ is the plastic strain in the state $\Gamma$.

Having discussed the notion of $ \textrm W_{el}({\bf{E}},\Gamma)$ and ${\mathrm W}_{nel}(\Gamma)$ in detail, we now would like to turn our attention to the essential attributes that define $\Gamma$ in materials undergoing progressive degradation. It should be recognized that the  material state $({\bf{E}},\Gamma)$ in this case evolves with time i.e., we have $({\bf{E}}(t),\Gamma(t))$. At any time `$t$', the state $({\bf{0}},\Gamma(t))$ represents the elastically unloaded configuration. This notion is akin to that of multiple natural configurations as introduced in \citet{rajagopal2004thermomechanics}. A complete characterization of material response involves determining how $\Gamma(t)$ evolves in time and is what we refer to as prognosis. Two important aspects that characterize $\Gamma(t)$ in a purely mechanical setting (no coupled process like chemo-mechanics) are geometry and microstructure of the material in the state $({\bf{0}},\Gamma(t))$. While the geometry determines how the density of the material changes due to the change in the $\Gamma$, microstructure characterizes the material response from that state. Imagine a sphere in the virgin state transformed to an ellipsoid in a different state $\Gamma$ with an accompanying change in volume. Balance of mass requires that the density be different in the two states. Likewise, a material that is isotropic in its virgin state $\Gamma_0$ may behave as an anisotropic material in a different state $\Gamma$ due to the microstructural changes it underwent. This needs to be taken into account in the elastic response of the material i.e., $ \textrm W_{el}({\bf{E}},\Gamma)$. The choice of internal variables $\Gamma$ actually depends on the physical mechanisms underpinning the damage progression. However, in many cases, experimental data drives constitutive model development and hence the choice of $\Gamma$. In the next section, we discuss the choice of constitutive response functionals $ \textrm W_{el}({\bf{E}},\Gamma)$ and $ \textrm W_{nel}(\Gamma)$. In particular, we consider the cases for which experimental data pertaining to nonlinear ultrasonic response (second harmonic generation) of materials is available.

\subsection{Choice of constitutive response functionals -- $ \mathrm W_{el}({\bf{E}},\Gamma)$ and ${\mathrm W}_{nel}(\Gamma)$  \label{sec2.3}}
In this section, we discuss the choice of $ \mathrm W_{el}({\bf{E}},\Gamma)$ and ${\mathrm W}_{nel}(\Gamma)$ for specific cases of damage progression delineating the physical motivation and experimental background pertaining to the choice.  We consider the cases for which it is assumed that the material state $\Gamma$ can be described by just a single internal variable i.e., $\Gamma=\{\Gamma_1\} $ and for brevity we drop the subscript. Moreover, we assume that $\Gamma \in [0,1]$ where $\Gamma=0$ represents the virgin or undamaged state and $\Gamma=1$ represents a completely damaged state --- a commonly adopted notion in damage mechanics. First, we discuss the nonlinear elastic response function of the material ($ \mathrm W_{el}({\bf{E}},\Gamma)$) and then followed by ${\mathrm W}_{nel}(\Gamma)$.

\subsubsection{$\mathrm W_{el}({\bf{E}},\Gamma) $}
Since our main goal is to develop a framework intended for damage sensing and prognosis using nonlinear ultrasonic techniques especially second harmonic generation, we start with the widely used Landau-Lifshitz model \citep{landau1986theory} (Eqn(\ref{eqn7})) to represent the isotropic hyperelastic material response $\textrm W_{el}({\bf{E}},\Gamma)$ from each state $\Gamma$ i.e., 
\begin{equation} \label{eqn7}
	\textrm W_{el}({\bf{E}},\Gamma)=\frac{1}{2} \lambda (\Gamma) (tr({\bf{E}}))^2+ \mu (\Gamma) tr({\bf{E}}^2) +\frac{1}{3} \mathrm C(\Gamma) (tr({\bf{E}}))^3 + \mathrm B(\Gamma) tr({\bf{E}})tr({\bf{E}}^2)+\frac{1}{3} \mathrm A(\Gamma) tr({\bf{E}}^3) 
\end{equation}   
Here, $\lambda (\Gamma)$ and $\mu (\Gamma)$	denote the lame's constants and $( \mathrm A(\Gamma),\mathrm B(\Gamma),\mathrm C(\Gamma))$ denote the third order elastic constants. Note that the above choice of constitutive model is suited to model second harmonic generation from classical nonlinearity in materials. One may have to use a different constitutive model for either modeling third harmonic generation \citep{chillara2016constitutive} or a non-classical nonlinear response.

 From Eqn(\ref{eqn7}), it follows that characterizing $ \mathrm W_{el}({\bf{E}},\Gamma)$ involves determining the constants $\left(\lambda (\Gamma),\mu (\Gamma), \mathrm A(\Gamma), \mathrm B(\Gamma),\mathrm C(\Gamma)\right)$. In general, $\lambda(\Gamma)$ and $\mu(\Gamma)$ are assumed to be linear in $\Gamma$ i.e., $\lambda(\Gamma)=\lambda_0(1-a\Gamma)$ and $\mu(\Gamma)=\mu_0(1-a\Gamma)$ for some constant $0 \leq a\leq 1$. Our main interest is in developing models where  $\left(\mathrm A(\Gamma),\mathrm B(\Gamma),\mathrm C(\Gamma)\right)$ correspond to specific cases of damage progression as in fatigue, creep, plastic deformation, etc. It should be noted that while the linear elastic material properties, namely $\lambda(\Gamma)$ and $\mu(\Gamma)$ are generally assumed and also experimentally observed to monotonically decrease with damage/degradation, the nonlinear elastic properties $\left(\mathrm A(\Gamma),\mathrm B(\Gamma),\mathrm C(\Gamma)\right)$ are observed to obey monotonically increasing (fatigue \citep{cantrell2001nonlinear}) or non-monotonic trends (creep \citep{baby2008creep}) depending on the nature of degradation. So, to address this issue, we discuss some plausible choices for the nonlinear elastic material properties that show the above mentioned trends. Moreover, we assume the functional forms for $\mathrm A(\Gamma)$, $\mathrm B(\Gamma)$ and $\mathrm C(\Gamma)$ are identical i.e., $ \frac{\mathrm A(\Gamma)}{\mathrm A_0}=\frac{\mathrm B(\Gamma)}{\mathrm B_0}=\frac{\mathrm C(\Gamma)}{\mathrm C_0}$ where $\mathrm A(\Gamma=0)=\mathrm A_0$ and so on for $\mathrm B_0$ and $\mathrm C_0$.
 
 \subsection*{Monotonic nonlinear response with damage} 
\begin{figure}	
  	\centering
  	\includegraphics[scale=0.5]{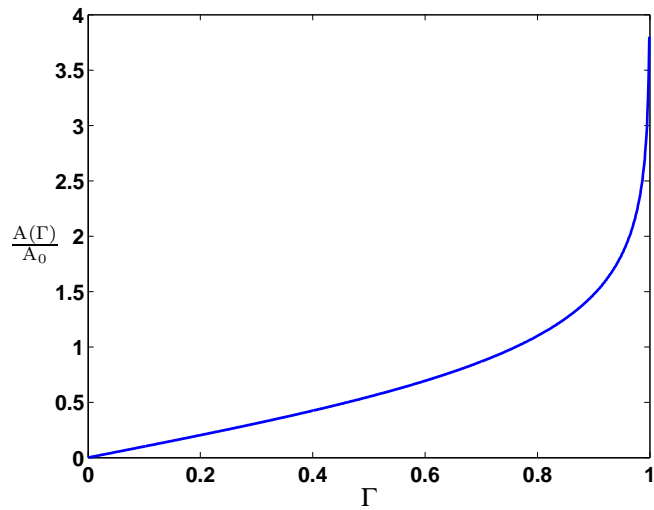}
  	\caption{A candidate function $\mathrm A(\Gamma)=\mathrm A_0 \tanh^{-1}(\Gamma)$ showing increasing nonlinearity parameter with increasing $\Gamma$. \label{fig2}}
  \end{figure}

\begin{figure}	
 	\centering
 	\includegraphics[scale=0.5]{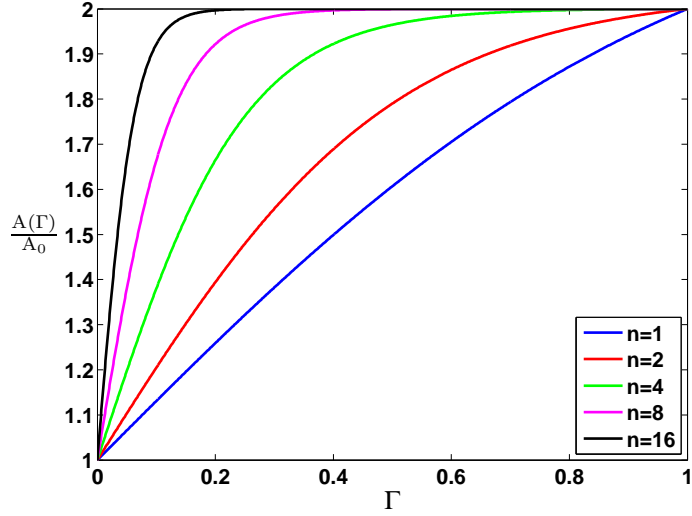}
 	\caption{A candidate function $\mathrm A(\Gamma)=\mathrm A_0(1+(m-1)\tanh(n\Gamma)/\tanh(n))$ with $m=2$ showing increasing nonlinearity parameter that asymptotically approaches $2 \mathrm A_0$ for different $n$. \label{fig3}}
 \end{figure}

 Figure \ref{fig2} shows a candidate function $\mathrm A(\Gamma)$ where the nonlinearity parameter is monotonically increasing with damage i.e., $\Gamma$. Here we chose
 \begin{equation} \label{eqn7a}
 \mathrm A(\Gamma)=\mathrm A_0\tanh^{-1}(\Gamma).
 \end{equation}
As $\Gamma \rightarrow 1$, we have $\mathrm A(\Gamma) \rightarrow \infty$ and $\frac{\partial \mathrm A(\Gamma)}{\partial \Gamma} \rightarrow \infty$ i.e., both the nonlinearity and the rate of increase of nonlinearity go unbounded as $\Gamma$ approaches 1. On the other hand, Figure \ref{fig3} shows the plot for the function 
\begin{equation} \label{eqn7b}
\mathrm A(\Gamma)=\mathrm A_0 (1+(m-1)\frac{\tanh(n\Gamma)}{tanh(n)})
\end{equation}
 with $m=2$ for different values of $n$. The above choice of the function depicts a monotonically increasing nonlinearity that asymptotically approaches $m \mathrm A_0$ i.e., the nonlinearity in the damaged state is $m$ times its value in the virgin state. Such an asymptotic response was observed for the low-cycle fatigue behavior as discussed in \citet{pruell2009evaluation} (See Figure 7 in \citet{pruell2009evaluation}). Moreover, the behavior depicted in Figure \ref{fig3} for low values of $n$ is representative of fatigue-induced nonlinearity as discussed in \citet{cantrell2001nonlinear,cantrell2004substructural} (See Figure 2 in \citet{cantrell2001nonlinear} and Figure 1 in \citet{cantrell2004substructural}). 
 
 \subsection*{Non-monotonic nonlinear response with damage}               
\begin{figure}	
  	\centering
  	\includegraphics[scale=0.5]{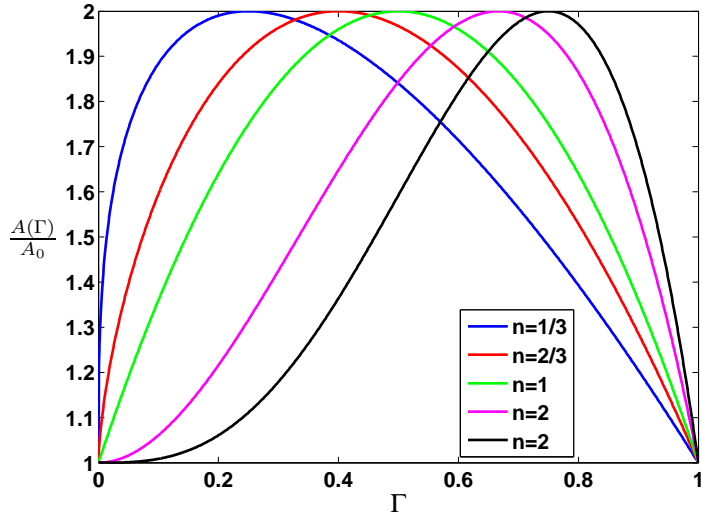}
  	\caption{ A candidate function $\mathrm A(\Gamma)=\mathrm A_0(1+\Gamma^n(1-\Gamma)(\frac{(n+1)^{(n+1)}}{n^n}))$ for nonlinearity parameter depicting a non-monotonic response with damage. \label{fig4}}
  \end{figure}

\begin{figure}	
 	\centering
 	\includegraphics[scale=0.5]{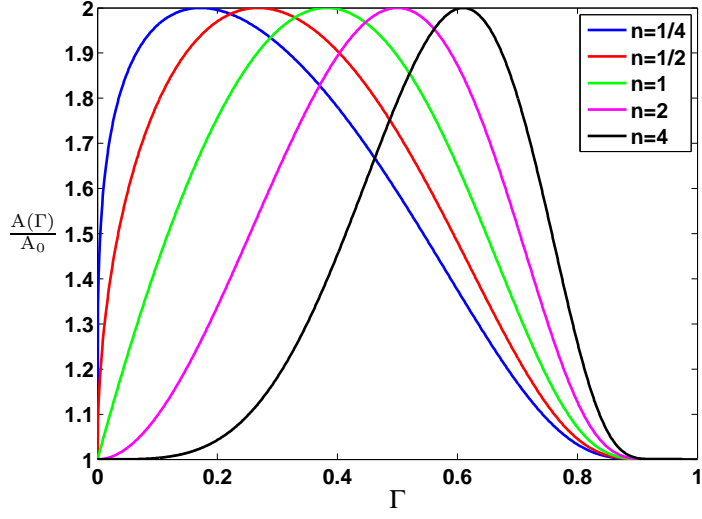}
 	\caption{A candidate function $\mathrm A(\Gamma)=\mathrm A_0(1+c\Gamma^n e^ {({-\frac{1}{1-\Gamma}})})$ for nonlinearity parameter depicting a non-monotonic response with damage (c is chosen so that $\max \{\mathrm A(\Gamma)\} =2 \mathrm A_0$ ). \label{fig5}}
 \end{figure}

As mentioned earlier, the nonlinearity parameter, in some cases may show a non-monotonic trend with increasing damage. This is the case for creep damage progression as discussed in \citet{baby2008creep} and \citet{xiang2014creep} (See Figure 5(c) in \citet{baby2008creep} and Figure 11 in \citet{xiang2014creep}). Figure \ref{fig4} shows a candidate function that is suitable to model the aforementioned non-monotonic nonlinear response. Here, we chose 
\begin{equation} \label{eqn7c}
\mathrm A(\Gamma)=\mathrm A_0(1+\Gamma^n(1-\Gamma)(\frac{(n+1)^{(n+1)}}{n^n})).
\end{equation}
As can be seen, the nonlinearity parameter increases, reaches a maximum at $\Gamma=\frac{n}{n+1}$ and then decreases. Likewise, Figure \ref{fig5} shows another candidate function 
\begin{equation} \label{eqn7d}
 \mathrm A(\Gamma)=\mathrm A_0(1+c\Gamma^n e^ {({-\frac{1}{1-\Gamma}})})
\end{equation}
which again shows a non-monotonic response. The constant $c$ is chosen so that $\mathrm A(\Gamma)$ attains a maximum value of $2 \mathrm A_0$. For $n<1$, both the functions show unbounded rate of increase at $\Gamma=0$. Also, for $n>1$, while the candidate function shown in Figure \ref{fig4} shows a zero slope at $\Gamma=0$ and a non-zero slope at $\Gamma=1$, the candidate function shown in Figure \ref{fig5} shows a zero slope at $\Gamma=0$ and $\Gamma=1$. Based on the above observations, it appears that a candidate function belonging to the family of functions depicted in Figure \ref{fig5} with $n>1$ is the most suitable one for modeling the creep-induced nonlinear response in metals. 

\subsection*{Damage induced anisotropic nonlinear response}    
Here, we discuss a plausible scenario wherein the damage induced nonlinearity shows anisotropic response. While experimental investigations have not discussed this aspect, it is certainly possible that the nonlinear response is anisotropic in cases where there is a preferred orientation for damage progression. For example, persistent slip bands (PSB's) that have a preferred orientation come under this category. For a general case of anisotropy, we have (up to third order in ${\bf E}$)

\begin{equation} \label{eqn8}
\textrm W_{el}({\bf{E}},\Gamma)=\frac{1}{2!} \mathrm{ C_{ijkl}}(\Gamma) \mathrm {E_{ij} E_{kl}}+\frac{1}{3!} \mathrm {C_{ijklmn}(\Gamma) E_{ij} E_{kl} E_{mn}}
\end{equation} 
where $\mathrm{C_{ijkl}}$ are second order elastic constants, $\mathrm{C_{ijklmn}}$ are third order elastic constants and $\mathrm{E_{ij}}$ is Lagrangian strain in index notation. Here, we consider the case where an initially isotropic material shows a nonlinear transversely isotropic response as damage progresses. For this case, we have $\Gamma=\{\Gamma_1, \Gamma_2, {\bf a}\}$ where $\Gamma_1$ and $\Gamma_2$ are damage variables and `${\bf a}$' is the unit vector that corresponds to the direction of transverse isotropy. Also, we assume that the isotropic damage response is characterized by $\Gamma_1$ and the anisotropic damage response is characterized by $\Gamma_2$. Then, we have
\begin{eqnarray} \label{eqn9}
\nonumber \textrm W_{el}({\bf{E}},\Gamma_1,\Gamma_2,{\bf a})&=&\frac{1}{2} \lambda (\Gamma_1) (tr({\bf{E}}))^2+ \mu (\Gamma_1) tr({\bf{E}}^2) +\frac{1}{3} \mathrm C(\Gamma_1) (tr({\bf{E}}))^3 + \\
\nonumber & & \mathrm B(\Gamma_1) tr({\bf{E}})tr({\bf{E}}^2)+\frac{1}{3} \mathrm A(\Gamma_1) tr({\bf{E}}^3)+ \mathrm D_1(\Gamma_2)({\bf a.Ea})^3+\\
\nonumber & & \mathrm D_2(\Gamma_2)({\bf a.Ea})^2 (tr({\bf{E}}))+\mathrm D_3(\Gamma_2)({\bf a.Ea})tr({\bf{E}}^2)+\\
\nonumber & & \mathrm D_4(\Gamma_2)({\bf a.Ea}) (tr({\bf{E}}))^2+\mathrm D_5(\Gamma_2)({\bf a.Ea})({\bf a.E^2a})+\\
&& \mathrm D_6(\Gamma_2)({\bf a.E^2a})(tr({\bf{E}}))
\end{eqnarray}  
where $\mathrm D_i \;(i=1,2,\cdots 6)$ are higher order anisotropic elastic constants with $\mathrm D_i(0)=0 \;(i=1,2,\cdots 6)$. Note that the parameter $\Gamma_2$ and the unit vector ${\bf a}$ enter the expression only in the nonlinear part i.e., in terms of order three in ${\bf E}$. In essence, the linear elastic response of damaged material is isotropic and the nonlinear elastic response is anisotropic. Note that the above choice is a Taylor-series expansion about ${\bf E=0}$ written in terms of invariants for the transversely anisotropic material. Such an analytical form was recently employed to study important features of nonlinear wave propagation in composites \citep{zhao2016second}. A particularly interesting aspect of employing the above Eqn (\ref{eqn9}) is capturing the onset of anisotropy i.e., the instant at which $\Gamma_2$ tends to be non-zero during the course of damage progression. Also, the direction of anisotropy (${\bf a}$) depends on the kind of loading the material is subjected to. A simple criterion that one may use to identify this direction is to resort to the notion akin to that of critical resolved shear stress in crystal plasticity. Moreover, it should be recognized that the evolution (or at least onset) of $\Gamma_2$ is dependent on $\Gamma_1$. Hence, it is an intricately coupled problem and requires a detailed discussion in its own right. This is beyond the scope of this article and the purpose of this discussion was just to bring the attention of the readers to this interesting aspect of the problem. Next, we discuss the choice of constitutive response function for energy dissipation i.e., ${\mathrm W_{nel}(\Gamma)}$.     

\subsubsection{$\mathrm W_{nel}(\Gamma) $}
 As mentioned earlier, $\mathrm W_{nel}(\Gamma)$ corresponds to the non-recoverable energy that is either stored in the material or dissipated from the material and plays a key role in damage progression. For brevity, we refer to $\mathrm W_{nel}(\Gamma)$ as dissipation or energy dissipated even though it corresponds to the total energy that is either dissipated or stored in an inelastic fashion.  Intuitively, it appears that as the material degrades, it loses its ability to store recoverable energy. Likewise, damage progression is invariably accompanied by energy dissipation i.e., as $\Gamma$ increases we expect $\mathrm W_{nel}(\Gamma) $ to increase. In other words, we must have $\mathrm W_{nel}(\Gamma)$ to be a nondecreasing function of $\Gamma$ i.e., $\frac{\partial \mathrm W_{nel}(\Gamma)}{\partial \Gamma} \geq 0$. Without loss of generality, one can assume that $\mathrm W_{nel}(0)=0$ and hence we have $\mathrm W_{nel}(\Gamma)\geq 0$ for $\Gamma \in [0,1]$. Under the above restrictions, the following possibilities exist for the choice of  $\mathrm W_{nel}(\Gamma)$ i.e.,
 \begin{enumerate}
 	
 \item $\mathrm W_{nel}(\Gamma=1)$ is finite and $ \frac{\partial \mathrm W_{nel}(\Gamma)}{\partial \Gamma}$ is bounded in $\Gamma \in [0,1]$ i.e, $ \frac{\partial \mathrm W_{nel}(\Gamma)}{\partial \Gamma} < \infty$.
 \item $\mathrm W_{nel}(\Gamma=1)$ is finite and $ \frac{\partial \mathrm W_{nel}(\Gamma)}{\partial \Gamma}$ is unbounded in $\Gamma \in [0,1]$ i.e., $ \frac{\partial \mathrm W_{nel}(\Gamma)}{\partial \Gamma} \rightarrow \infty$ for some $\Gamma \in [0,1]$. If one assumes $\mathrm W_{nel}(\Gamma)$ to be sufficiently smooth, then one can only have $\frac{\partial \mathrm W_{nel}(\Gamma)}{\partial \Gamma} \rightarrow \infty$ as $\Gamma \rightarrow 1$ or $\Gamma \rightarrow 0$. We restrict to the case where  $\frac{\partial \mathrm W_{nel}(\Gamma)}{\partial \Gamma} \rightarrow \infty$ as $\Gamma \rightarrow 1$.
 \item $\mathrm W_{nel}(\Gamma) \rightarrow \infty$ as $\Gamma \rightarrow 1$.
 
 \end{enumerate}     
Next we discuss some possible candidates for $\mathrm W_{nel}(\Gamma)$ that belong to the above cases.

\subsection*{ \centering $\mathrm W_{nel}(\Gamma=1)$ is finite and $\frac{\partial\mathrm W_{nel}(\Gamma)}{\partial \Gamma}$ is bounded}
	
\begin{figure}	
  	\centering
  	\includegraphics[scale=0.5]{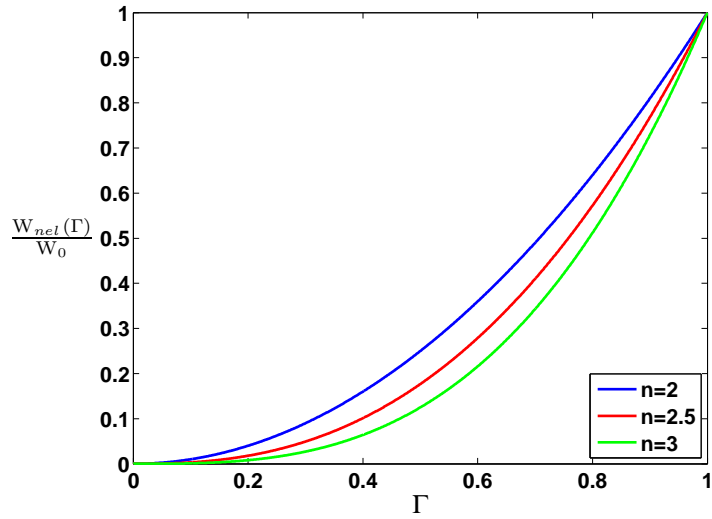}
  	\caption{A candidate function $\mathrm W_{nel}(\Gamma)=\mathrm W_0 \Gamma^n$ --- corresponds to finite $\mathrm W_{nel}(\Gamma=1)=\mathrm W_0$ and bounded $ \frac{\partial \mathrm W_{nel}(\Gamma)}{\partial \Gamma} < \infty$.  \label{fig6}}
  \end{figure}

$\mathrm W_{nel}(\Gamma=1)$ being finite corresponds to the scenario where the total energy dissipated during the entire damage progression  is finite.  Figure \ref{fig6} shows the plot of
\begin{equation} \label{eqn9a}
\mathrm W_{nel}(\Gamma)=\mathrm W_0 \Gamma^n
\end{equation}
for different $n>1$. Here, the total energy dissipated is finite and is equal to $\mathrm W_0$.
	
\subsection*{ \centering $\mathrm W_{nel}(\Gamma=1)$ is finite and $\frac{\partial\mathrm W_{nel}(\Gamma)}{\partial \Gamma}$ is unbounded}
\begin{figure}	
  	\centering
  	\includegraphics[scale=0.5]{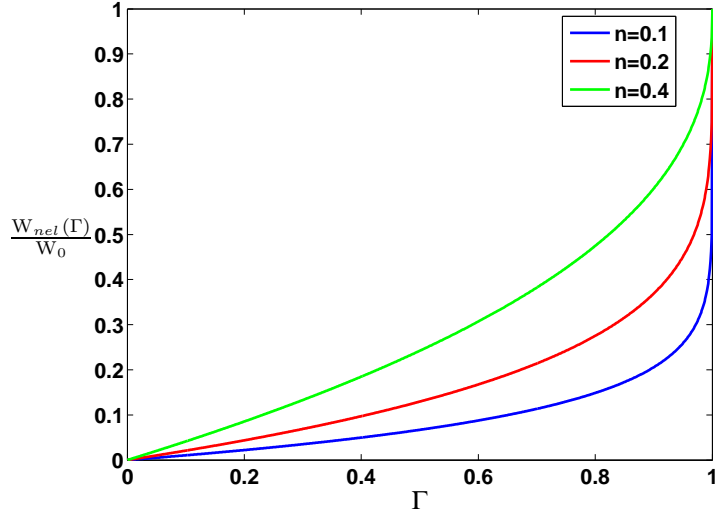}
  	\caption{A candidate function $\mathrm W_{nel}(\Gamma)=\mathrm W_0(1- (1-\Gamma)^n)$ --- corresponds to finite $\mathrm W_{nel}(\Gamma=1)=\mathrm W_0$ and unbounded $ \frac{\partial \mathrm W_{nel}(\Gamma)}{\partial \Gamma}$. \label{fig7}}
  \end{figure}

Figure \ref{fig7} shows the plot of 
\begin{equation}\label{eqn9b}
 \mathrm W_{nel}(\Gamma)=\mathrm W_0(1-(1-\Gamma)^n)
\end{equation}  
for $n<1$. Clearly, $W_{nel}(\Gamma=1)=\mathrm W_0$ is finite and $\frac{\partial\mathrm W_{nel}(\Gamma)}{\partial \Gamma}$ is unbounded as $\Gamma \rightarrow 1$. This case corresponds to the scenario where the total energy dissipated is finite ($\mathrm W_0$) but infinite driving force ($\frac{\partial\mathrm W_{nel}(\Gamma)}{\partial \Gamma}$)  is necessary for complete damage progression to occur.
 
\subsection*{ \centering $\mathrm W_{nel}(\Gamma) \rightarrow \infty$  and $\frac{\partial\mathrm W_{nel}(\Gamma)}{\partial \Gamma} \rightarrow \infty $ as $\Gamma \rightarrow 1$}
\begin{figure}	
	\centering
	\includegraphics[scale=0.5]{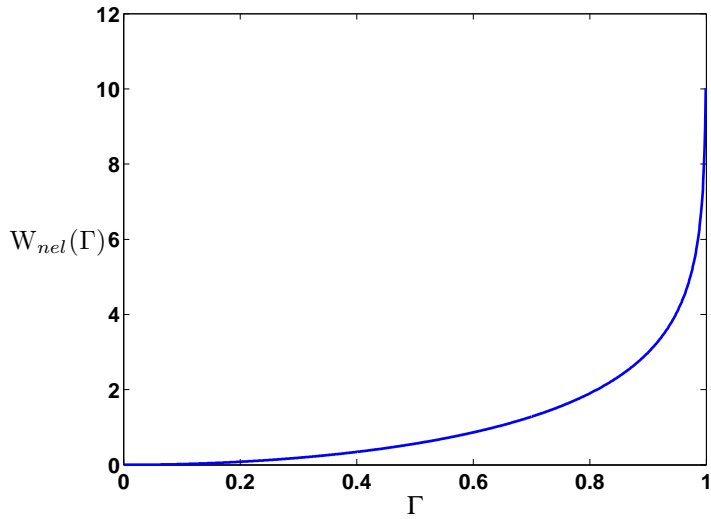}
	\caption{A candidate function $\mathrm W_{nel}(\Gamma)=(\Gamma\tanh^{-1}(\Gamma)-\log(1-\Gamma^2))$ --- corresponds to unbounded $\mathrm W_{nel}(\Gamma)$ and unbounded $ \frac{\partial \mathrm W_{nel}(\Gamma)}{\partial \Gamma}$. \label{fig8}}
\end{figure}

Figure \ref{fig8} shows the plot of 
\begin{equation} \label{eqn9c}
\mathrm W_{nel}(\Gamma)=(\Gamma\tanh^{-1}(\Gamma)-\log(1-\Gamma^2)). 
\end{equation}
For this case, infinite energy needs to be dissipated for complete damage progression to occur. In other words, $\Gamma$ can never be equal to one.     

\section{Material state, damage growth, and prognosis}
Prognosis methodologies attempt to predict the course of damage growth having known the current material state, damage-history and loads acting on the material. Generally, these methodologies are an outcome of large sets of experimental data where empirical relations are sought to describe the data based on experimental measurables. For example, Paris law and Palmgren-Miner rule for fatigue crack growth belong to this category. The above empirical relations are damage-specific and hence are applicable to specific cases they are intended for. Moreover, the empirical relations may not have a thermodynamic basis i.e., they may not consider the energetics or kinetics associated with damage growth. In other words, they are data-driven models as opposed to being physics-based.  However, the success of such empirical relations in predicting damage growth cannot be undermined. Another approach that is employed to study damage prognosis is continuum damage mechanics \citep{kachanov2013introduction,lemaitre1984use,chaboche1988continuum} where the damage is identified with internal variables and damage growth is regarded as a thermodynamic process. In this case, the evolution equations for damage are either prescribed empirically or are obtained as part of a general constitutive framework like that developed in \citet{rajagopal2007shear}. Herein, we develop a material-state-based damage prognosis approach in the context of thermodynamic framework introduced in section \ref{sec2}. Since the framework discussed earlier is independent of $\dot{\Gamma}$, we are considering the case where material deformation and damage growth is assumed to be quasi-static (See Remark \ref{rem3}). Such an assumption is valid when the damage growth rate is slow in that $\dot{\Gamma} \approx 0$. Also, while the above framework does not incorporate an explicit time-dependence, damage scenarios like fatigue can be modeled by considering the $\mathrm W_{nel}(\Gamma)$ to be dependent on the frequency ($f$) of the loading and number of fatigue cycles ($\mathrm N$) as additional parameters in addition to $\Gamma$. The outcome of the present approach is that it results in explicit relations between macroscopic material response and the internal damage variables. These are obtained by appealing to the variational principles for the thermodynamic system (the material undergoing progressive damage) governed by the pseudo-elastic strain energy function introduced in section \ref{sec2}. Through a set of examples we demonstrate the applicability of the framework for material-state determination and prognosis using nonlinear ultrasonics. 
\begin{remark}\label{rem3}
Our main interest here is in developing models for ultrasonic NDE and SHM. For these cases, the time-scales (a few milliseconds) associated with a single measurement is much smaller compared to those associated with damage growth. Hence the ``quasi-static'' assumption is very reasonable in this context.    	
\end{remark}

\subsection{Spring-mass-system}
In this section we employ the framework to model damage growth in a spring-mass system where the spring undergoes progressive degradation. Mechanical analogs incorporating spring and dash-pots are often employed to understand material behavior.
  
\begin{figure}	
 	\centering
    \psfrag{x1}{$x_0(\Gamma)$}
    \psfrag{x}{$x$}
    \psfrag{m}{$m$}
    \psfrag{F}{$\mathrm F$}
    \psfrag{k0}{$k(\Gamma)$}
 	\includegraphics[scale=1]{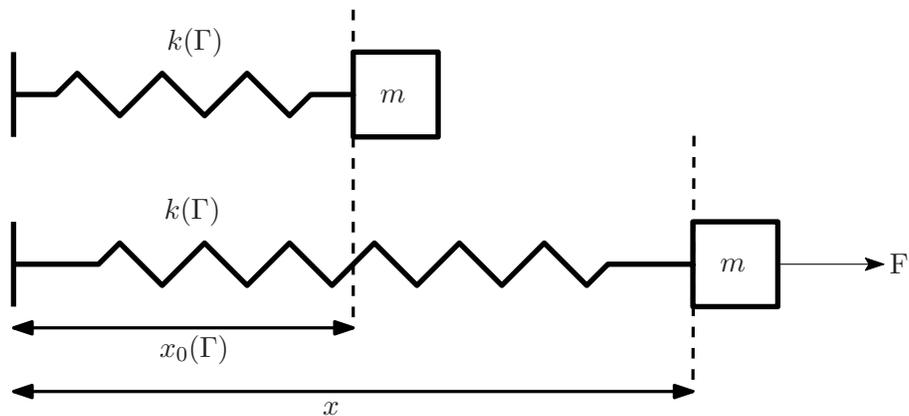}
 	 	\caption{Schematic of the spring-mass system with the spring undergoing degradation. \label{fig9}}
 	
 \end{figure}

Consider the schematic of a spring-mass system with a mass $m$ acted upon by a force $\mathrm F$ as shown in the Figure \ref{fig9}. The stiffness of the spring ($k(\Gamma)$)) and the unstretched length ($x_0(\Gamma)$) of the spring are assumed to be dependent on the damage state ($\Gamma$). For each damage state $\Gamma$, the system when unloaded returns itself to the state with mass at position $x=x_0(\Gamma)$. Damage growth manifests in two different ways
\begin{enumerate}
	\item Reduction in the stiffness of the spring.
	\item A permanent increase in the unstretched length ($x_0(\Gamma)$) of the spring that corresponds to plastic deformation. 
\end{enumerate}

\subsubsection{Stress relaxation-like response}
\begin{figure}	
 	\centering
 	\psfrag{x1}{$x_0(\Gamma)$}
 	\psfrag{x}{$x_m$}
 	\psfrag{m}{$m$}
 	\psfrag{F}{$\mathrm F(\Gamma)$}
 	\psfrag{k0}{$k(\Gamma)$}
 	\includegraphics[scale=1]{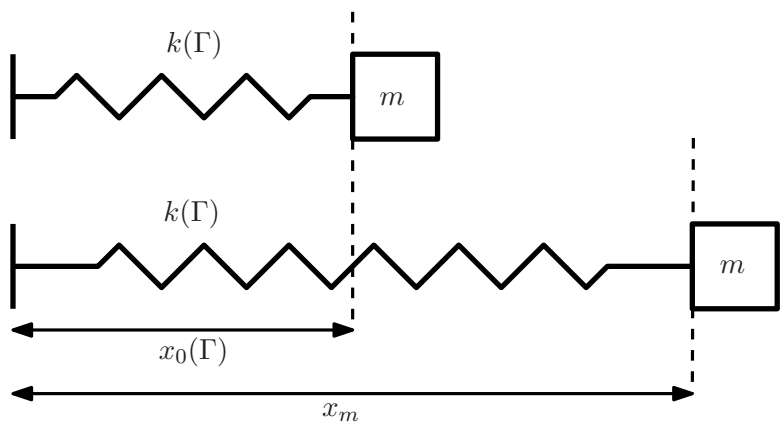}
 	\caption{Schematic of the spring mass system used to study the relaxation-like response. \label{fig10}}
 	
 \end{figure}

Now, we consider the spring-mass system shown in Figure \ref{fig10} to study the relaxation response of the spring undergoing degradation. Here, we assume that the mass is displaced to the position $x=x_m$ and is then held there as shown in the Figure \ref{fig10}. We then examine the response of the spring as it undergoes degradation i.e., we evaluate the force required to maintain the position of the mass at $x=x_m$. As the stiffness of the spring reduces with increasing damage, the force required to maintain the mass at $x=x_m$ continuously decreases and eventually goes to zero. At the same time, the unstretched length of the spring $x_0(\Gamma) \rightarrow x_m$ as $\Gamma \rightarrow 1$. For the present case, we have

\begin{equation} \label{eqn14}
\textrm W(x,\Gamma)=\textrm W_{el}(x,\Gamma)+\textrm W_{nel}(\Gamma)
\end{equation} 
with 
\begin{equation}\label{eqn15}
\mathrm W_{el}(x,\Gamma)=\frac{1}{2} k(\Gamma) (x_m-x_0(\Gamma))^2.
\end{equation} 
The governing equation of equilibrium is obtained by setting the variation $\delta \mathrm W(x,\Gamma)=0$ and is given by 

\begin{equation}\label{eqn16}
\frac{1}{2}\frac{\partial k(\Gamma)}{\partial \Gamma} (x_m-x_0(\Gamma))^2+k(\Gamma)(x_m-x_0(\Gamma)) (-\frac{\partial x_0(\Gamma)}{\partial \Gamma})+\frac{\partial \mathrm W_{nel} (x,\Gamma)}{\partial \Gamma}=0
\end{equation} 
 
The above equation is an Ordinary-Differential-Equation (ODE) to be solved for $x_0(\Gamma)$. For the present case, we assume $k(\Gamma)=k_0(1-\Gamma)$ and from Eqn(\ref{eqn16}) we get the following algebraic equation that needs to be solved for $x_0(\Gamma)$.

\begin{equation}\label{eqn17}
\frac{1}{2}k_0(1-\Gamma) (x_m-x_0(\Gamma))^2=\frac{1}{2}k_0(x_m-x_0(0))^2-\mathrm W_{nel}(\Gamma)
\end{equation}    
It is here that the choice of the function $\mathrm W_{nel}(\Gamma)$ comes into play. We demonstrate the relaxation behavior for two different choices of $\mathrm W_{nel}(\Gamma)$.

\subsection*{ Case 1}
\begin{figure}	
 	\centering
 	\includegraphics[scale=0.5]{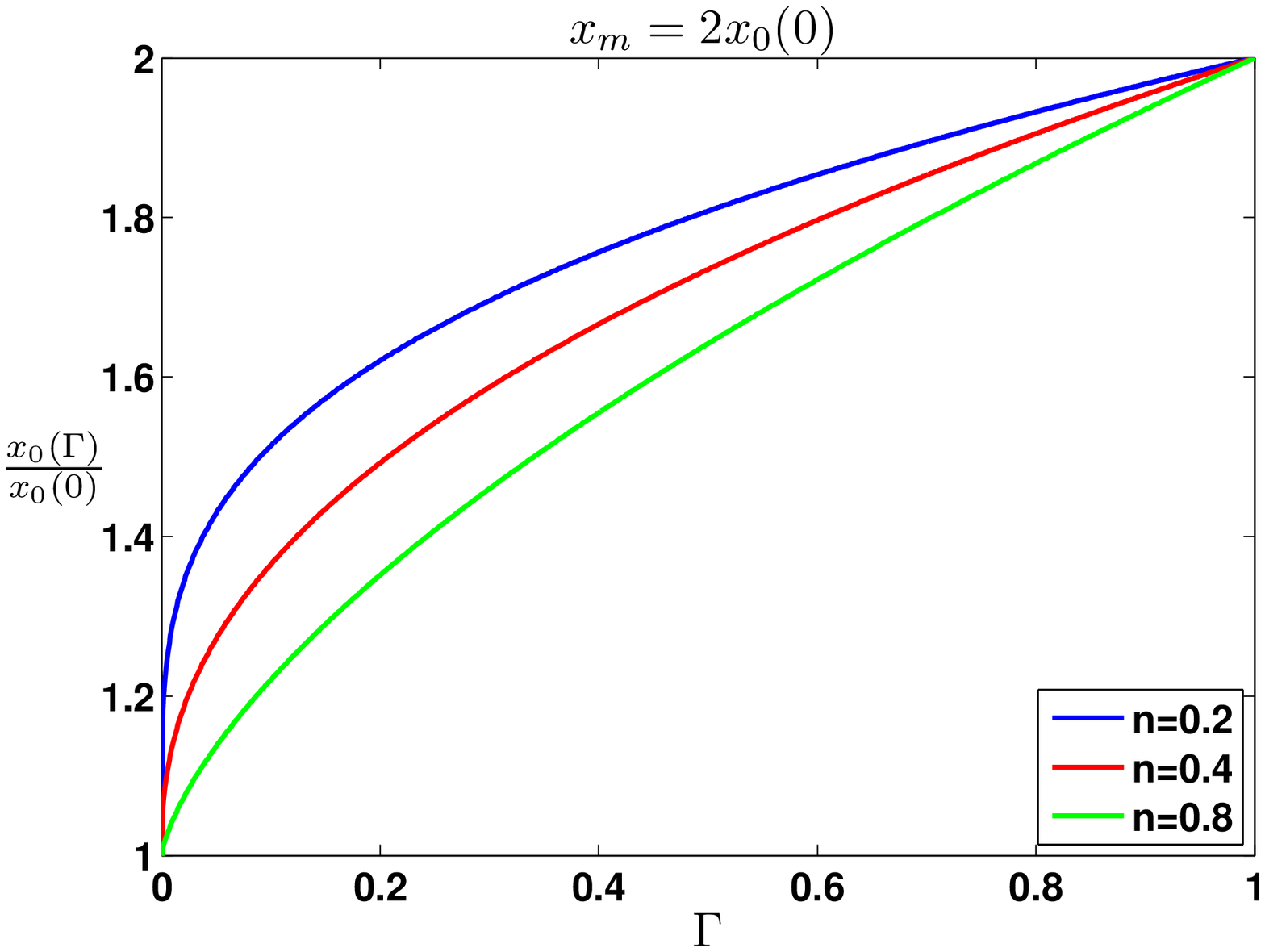}
 	\caption{Normalized unstretched length of the spring $\frac{x_0(\Gamma)}{x_0(0)}$ as a function of $\Gamma$ with $x_m=2x_0(0)$ and $\mathrm W_{nel}(\Gamma)=\mathrm W_0 \left (\frac{\Gamma^n-n\Gamma} {1-n}\right)$. \label{fig11}}
 \end{figure}

\begin{figure}	
	\centering
	\includegraphics[scale=0.5]{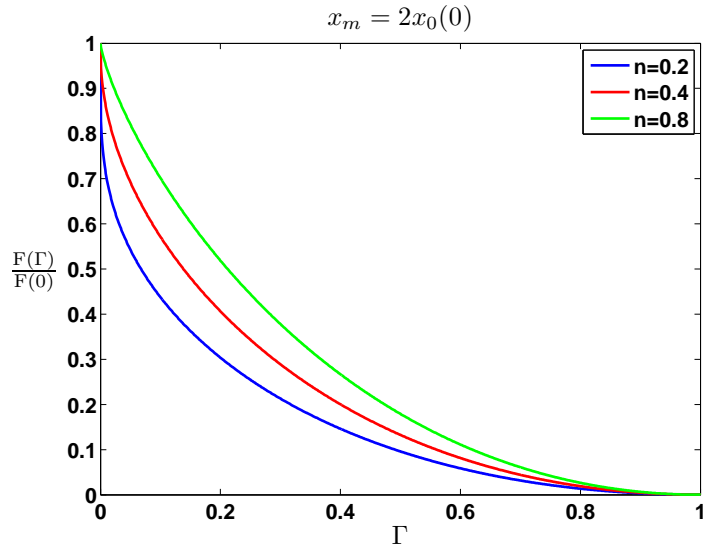}
	\caption{Normalized force in the spring $\frac{\mathrm F(\Gamma)}{\mathrm F(0)}$ as a function of $\Gamma$ with $x_m=2x_0(0)$ and $\mathrm W_{nel}(\Gamma)=\mathrm W_0 \left (\frac{\Gamma^n-n\Gamma} {1-n}\right)$. \label{fig12}}
\end{figure}

We choose 
$$\mathrm W_{nel}(\Gamma)=\mathrm W_0 \left (\frac{\Gamma^n-n\Gamma} {1-n}\right)$$ with $n<1$ and $$\mathrm W_0 =\frac{1}{2}k_0(x_m-x_0(0))^2.$$
Note that the above choice implies that the total energy dissipated during the relaxation process i.e., $\mathrm W_{nel} (\Gamma=1)=\mathrm W_0=\frac{1}{2}k_0(x_m-x_0(0))^2$ and is equal to the initial energy stored in the spring when $\Gamma=0$. This is consistent with the thermodynamics of the physical process at hand. Moreover, for the above choice of $\mathrm W_{nel}(\Gamma)$, one can easily solve Eqn(\ref{eqn17}) to obtain 
\begin{equation} \label{eqn18}
x_0(\Gamma)=\left(x_m-\sqrt{1-\frac{\Gamma^n-n\Gamma}{1-n}}(x_m-x_0(0))\right).
\end{equation}
Figure \ref{fig11} shows the normalized unstretched length of the spring $ \frac{x_0(\Gamma)}{x_0(0)}$ as a function of damage for $x_m=2x_0(0)$. As can be seen, the unstretched length of the spring monotonically increases and finally attains the value $x_0(\Gamma=1)=x_m=2x_0(0)$ where the spring is completely relaxed i.e., unstretched and the degradation does not proceed further. For this case, the force in the spring as a function of $\Gamma$ is given by $\mathrm F(\Gamma)=k(\Gamma)(x_m-x_0(\Gamma))$. Figure \ref{fig12} shows the normalized force $\frac{\mathrm F(\Gamma)}{\mathrm F(0)}$ as a function of damage. Clearly, the force in the spring goes to zero as $\Gamma \rightarrow 1$ showing the relaxation behavior. Note that $\Gamma$ attains the value 1 and degradation stops. Next, we consider the case where $\Gamma\rightarrow 1$ asymptotically, that is a more realistic description of relaxation response.     
 
\subsection*{ Case 2}
\begin{figure}	
 	\centering
 	\includegraphics[scale=0.5]{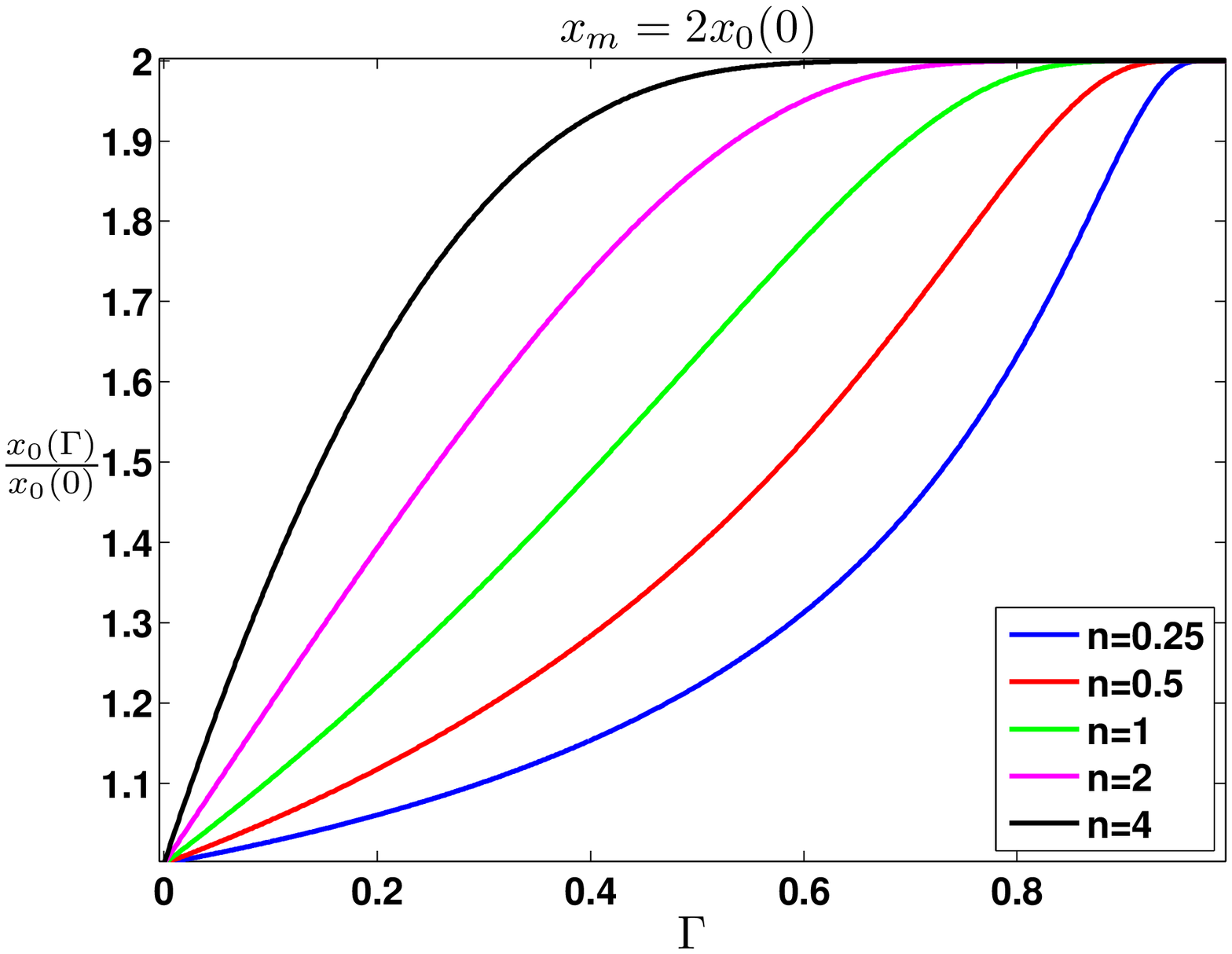}
 	\caption{Normalized unstretched length of the spring $\frac{x_0(\Gamma)}{x_0(0)}$ as a function of $\Gamma$ with $x_m=2x_0(0)$ and $\mathrm W_{nel}(\Gamma)=\mathrm W_0 (1-e^{\frac{-n\Gamma}{1-\Gamma}})$. \label{fig13}}
 \end{figure}

\begin{figure}	
 	\centering
 	\includegraphics[scale=0.5]{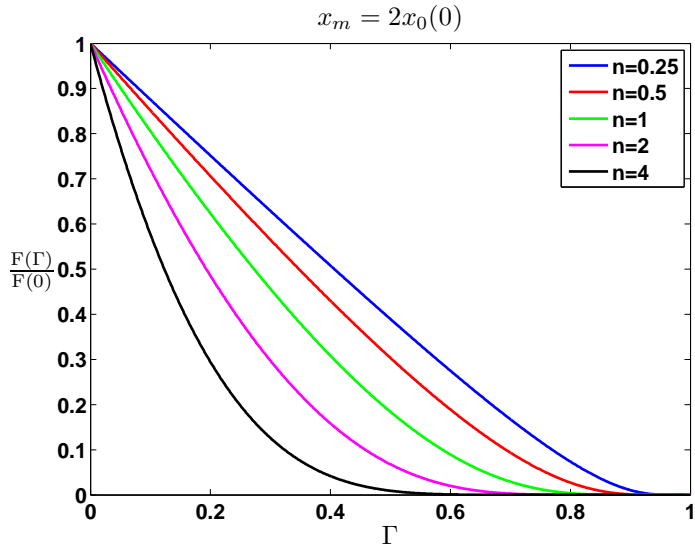}
 	\caption{Normalized force in the spring $\frac{\mathrm F(\Gamma)}{\mathrm F(0)}$ as a function of $\Gamma$ with $x_m=2x_0(0)$ and $\mathrm W_{nel}(\Gamma)=\mathrm W_0(1-e^{\frac{-n\Gamma}{1-\Gamma}})$. \label{fig14}}
 \end{figure}

Here, we choose 
$$\mathrm W_{nel}(\Gamma)=\mathrm W_0(1-e^{\frac{-n\Gamma}{1-\Gamma}})$$ with $n>1$ and $$\mathrm W_0 =\frac{1}{2}k_0(x_m-x_0(0))^2.$$
For the above choice, from Eqn(\ref{eqn17}), we get  
\begin{equation} \label{eqn20}
x_0(\Gamma)=\left(x_m-e^{-\frac{n\Gamma}{1-\Gamma}}(x_m-x_0(0))\right)
\end{equation}
Figure \ref{fig13} shows the normalized unstretched length of the spring $\frac{x_0(\Gamma)}{x_0(0)}$ as a function of damage for $x_m=2x_0(0)$ for different values of $n$. As can be seen, the unstretched length of the spring monotonically increases and approaches the value $x_0(\Gamma=1)=x_m=2x_0(0)$ asymptotically. Here $\Gamma=1$ is not attained and hence degradation is never complete. This asymptotic response is a more realistic depiction of relaxation. Also, Figure \ref{fig14} shows the normalized force $\frac{\mathrm F(\Gamma)}{\mathrm F(0)}$ as a function of damage. Clearly, the force in the spring asymptotically goes to zero as $\Gamma \rightarrow 1$ showing the relaxation behavior. It should be noted that the above treatment can be easily extended to a material undergoing degradation by replacing the stiffness of the spring with the Young's modulus.

\subsubsection*{Modeling the nonlinear response of the spring undergoing degradation}
\begin{figure}	
 	\centering
 	\includegraphics[scale=0.5]{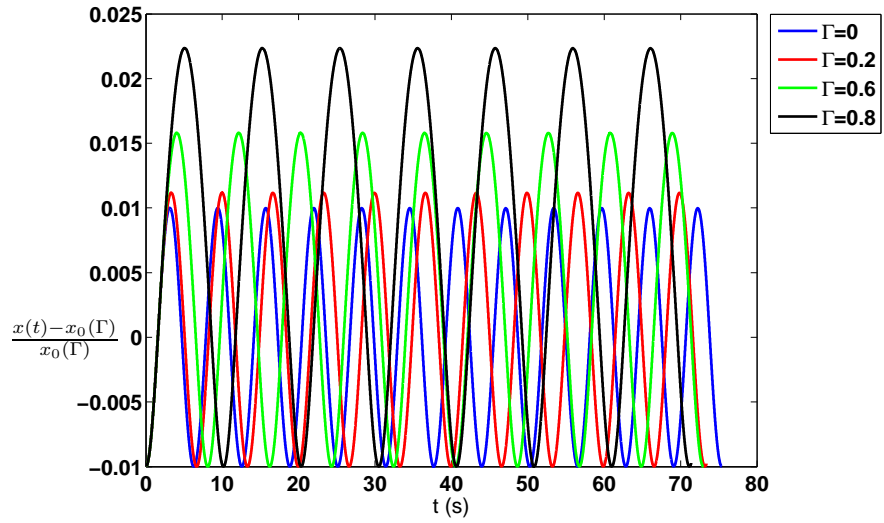}
 	\caption{Time-domain displacement response of the mass for different $\Gamma$.  \label{fig15}}
 \end{figure}

\begin{figure}	
 	\centering
 	\includegraphics[scale=0.5]{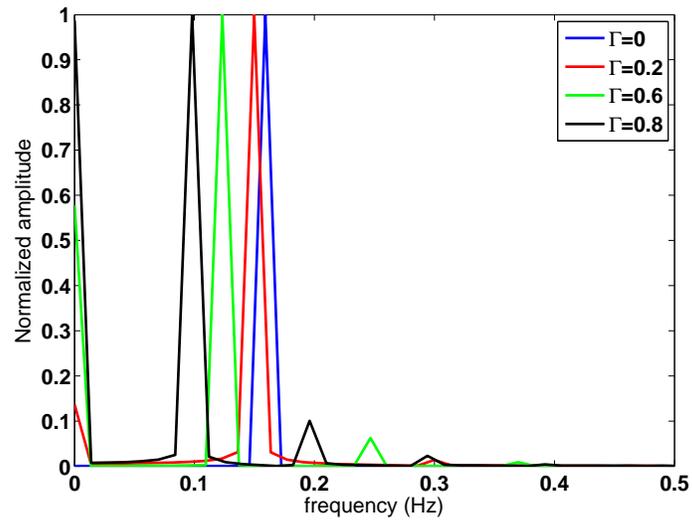}
 	\caption{Normalized FFT of the displacement response for different $\Gamma$.  \label{fig16}}
 \end{figure}

\begin{figure}
  	\centering
  	\includegraphics[scale=0.5]{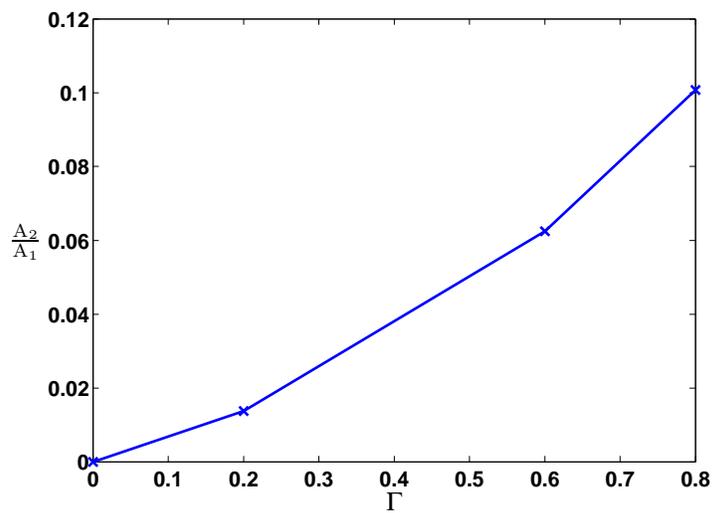}
  	\caption{Normalized second harmonic amplitude with $\Gamma$ . \label{fig17}}
  \end{figure}

  Now, we consider the nonlinear response of the spring undergoing degradation. To that end, we assume that the response of the spring at each damage state ($\Gamma$) is bilinear with stiffness $k(\Gamma)$ given by
  
  \begin{equation} \label{eqn21}
  k(\Gamma)=\begin{cases}
  k_0(1-\Gamma) & \mathrm{in\,tension} \, (x>x_0(\Gamma))\\
  k_0 & \mathrm{in\,compression} \, (x<x_0(\Gamma))
  \end{cases}
  \end{equation}
 where $k_0$ is the stiffness of the spring in its undamaged state $\Gamma=0$. Here, we are assuming that the stiffness of the spring in tension is decreasing as the spring degrades while the stiffness in compression in unaltered. It is a non-classical nonlinear response. Suppose that one has necessary instrumentation to measure the small-amplitude vibration response of the spring-mass system about the unloaded state $x=x_0(\Gamma)$. It should be expected that the vibration response of the mass shows higher harmonic generation due to the nonlinear response of the spring. Our interest is in examining how this nonlinear response changes with increasing $\Gamma$. For this study, we assume that the mass is displaced to $x=0.99x_0(\Gamma)$ and is then allowed to oscillate. These small-amplitude vibrations are assumed to not result in significant change in $\Gamma$. Figure \ref{fig15} shows the time domain response of the mass undergoing oscillations for different damage states $\Gamma=0 ,\,0.2,\, 0.6 ,\, \mathrm{and} \,0.8$. Clearly, the time-domain response tends to become asymmetric with increasing damage and hence generates second harmonics as discussed in \citet{chillara2014ijes}. Figure \ref{fig16} shows the FFT (Fast Fourier Transform) of the time domain responses shown in Figure \ref{fig15}. As can be seen, for non-zero $\Gamma$, we clearly have second harmonic in FFT's that increase  with $\Gamma$. Figure \ref{fig17} shows the normalized amplitude of second harmonic with increasing $\Gamma$. Here the amplitude $A_2$ of the second harmonic is normalized with amplitude $A_1$ of the first harmonic. Note that the increase of second harmonic amplitude is nonlinear with $\Gamma$. In the next section, we model the creep-like degradation of a material under the framework introduced in section \ref{sec2.2}

 \subsection{Creep-like degradation}
 In this section, we investigate the degradation behavior of a material showing a creep-like response under the constitutive framework introduced in section \ref{sec2.2}. To keep it simple, we discuss this in a 1D context under a linearized strain assumption. Consider a 1D bar loaded  under a constant uniaxial stress $\sigma$ and suppose that its Young's modulus in the virgin state is given by $\mathrm E_0$. For each material state $\Gamma$, we denote the plastic strain accumulated in the material by $\epsilon_p(\Gamma)$. As the material degrades, we expect $\epsilon_p(\Gamma)$ to monotonically increase with $\Gamma$ depicting a creep-like behavior. 
 
 We start with the pseudo-elastic strain energy function given by 
 
  \begin{equation} \label{eqn22}
  \textrm W(\epsilon,\Gamma)=\textrm W_{el}(\epsilon,\Gamma)+\textrm W_{nel}(\Gamma)
  \end{equation} 
where $\epsilon$ is the total strain. In addition, we choose the following form for $\textrm W_{el}(\epsilon,\Gamma)$ 
  
  \begin{equation} \label{eqn23}
  \textrm W_{el}(\epsilon,\Gamma)=\frac{1}{2} \mathrm E(\Gamma) (\epsilon - \epsilon_p(\Gamma))^2 + \frac{1}{3} \mathrm A(\Gamma) (\epsilon - \epsilon_p(\Gamma))^3
  \end{equation} 
  where $\mathrm E(\Gamma)$ and $\mathrm A(\Gamma)$ denote the Young's modulus and higher order elastic constant respectively. Under a quasi-static assumption on damage growth, one can obtain the governing equations by considering $\delta \textrm W(\epsilon,\Gamma)-\sigma.\epsilon=0$ which gives      
 
  \begin{subequations} 
  \begin{eqnarray}
  \frac{\partial \mathrm W_{el} (\epsilon,\Gamma)}{\partial \epsilon}-\sigma&=& 0 \label{eqn24a}\\
  \frac{\partial \mathrm W_{el} (\epsilon,\Gamma)}{\partial \Gamma}+ \frac{\partial \mathrm W_{nel} (\Gamma)}{\partial \Gamma}&=& 0. \label{eqn24b}
  \end{eqnarray}
  \end{subequations}  
 For the choice of $\textrm W_{el}(\epsilon,\Gamma)$ in Eqn(\ref{eqn23}), we get 
\begin{subequations} 
	\begin{eqnarray}
	\mathrm E(\Gamma) (\epsilon-\epsilon_p)+\mathrm A(\Gamma) (\epsilon-\epsilon_p)^2-\sigma&=& 0 \label{eqn25a}\\
	\nonumber \frac{\partial \mathrm W_{nel} (\Gamma)}{\partial \Gamma}+\frac{1}{2} \frac { \partial \mathrm E(\Gamma)}{\partial \Gamma} (\epsilon - \epsilon_p(\Gamma))^2+\frac{1}{3} \frac { \partial \mathrm A(\Gamma)}{\partial \Gamma} (\epsilon - \epsilon_p(\Gamma))^3\\-(\mathrm E(\Gamma) (\epsilon-\epsilon_p)+\mathrm A(\Gamma) (\epsilon-\epsilon_p)^2) \frac{\partial \epsilon_p}{\partial \Gamma}&=& 0. \label{eqn25b}
	\end{eqnarray}
\end{subequations}  	
Eqn (\ref{eqn25a}) and Eqn (\ref{eqn25b}) need to be solved for each $\Gamma$. We choose $\mathrm E(\Gamma)=\mathrm E_0(1-a\Gamma)$ with $a=0.5$ and $\mathrm E_0=70 \; \mathrm {GPa}$. Eqns (\ref{eqn25a} \& \ref{eqn25b}) together have three unknown functions of $\Gamma$--- $\mathrm W_{nel}(\Gamma)$, $\mathrm A(\Gamma)$, and $\epsilon_p(\Gamma)$. One needs to assume functional forms for any two of the above functions and solve for the other. Here, we choose  $\mathrm A(\Gamma)$ and $\mathrm W_{nel}(\Gamma)$. There is considerable evidence that the nonlinearity parameter during the creep damage growth follows a non-monotonic trend as in Figure \ref{fig5}. So, we choose, $\mathrm A(\Gamma)=\mathrm A_0(1+c\Gamma^n e^{(-\frac{1}{1-\Gamma})})$ with `$c$' chosen such that $\max \{\mathrm A(\Gamma)\}=2 \mathrm A_0$. In addition, we choose $\mathrm W_{nel}(\Gamma)=\mathrm K\Gamma+ \mathrm W_0 \Gamma^m $. This choice ensures that the total energy dissipated during the damage growth is $\mathrm K +\mathrm W_0$ and is independent of `$m$'. Moreover, the first term in $\mathrm W_{nel}(\Gamma)$ i.e., $\mathrm K\Gamma$ corresponds to dissipation with a constant driving force `$\mathrm K$'. The second term corresponds to a damage-dependent driving force and is governed by the exponent `$m$'. For large `$m$', initial damage growth is dominated by the first term  and its growth towards the end is dominated by the second term. Once the above choices are made, one can solve Eqn (\ref{eqn25a}) and Eqn (\ref{eqn25b}) to obtain the creep-strain $\epsilon_p(\Gamma)$. We present results for two choices of $\mathrm K$ and  $\mathrm W_0$.

\subsubsection*{Case 1: $\mathrm K= \mathrm W_0= 2 \mathrm {\,MPa} ; \; \sigma=140 \; \mathrm{MPa} $}
\begin{figure}
  	\centering
  	\includegraphics[scale=0.5]{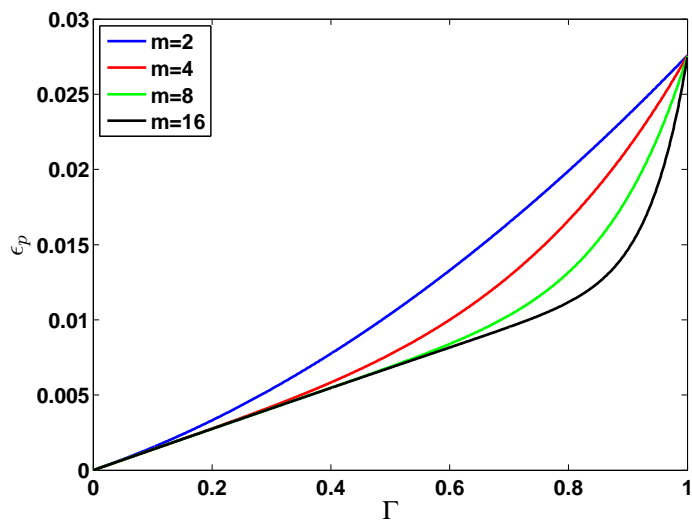}
  	\caption{Creep strain with $\Gamma$ for $m=2\,, 4\,, 8\, ,16$ . \label{fig18}}
  \end{figure}

\begin{figure}[h]
  	\centering
  	\includegraphics[scale=0.5]{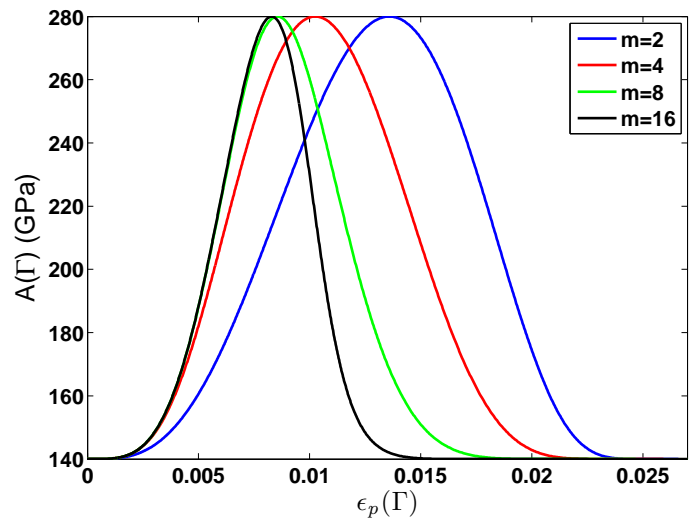}
  	\caption{Nonlinearity parameter $\mathrm A(\Gamma)$ with creep strain for $m=2\,, 4\,, 8\, ,16$. \label{fig19}}
  \end{figure}

Here, we assume that $\mathrm W_{nel}(\Gamma)$ is independent of stress ($\sigma$). Figure \ref{fig18} shows the creep-strain with $\Gamma$ for different `$m$'. As can be seen, for $m>2$, the creep-strain during the initial-phase of damage growth is independent of $m$. However, they differ considerably at higher $\Gamma$. Figure \ref{fig19} shows the nonlinearity parameter with creep-strain for different `$m$'. Clearly, it shows a non-monotonic trend with the creep-strain.

\subsubsection*{Case 2: $\mathrm K= \mathrm W_0= \sigma^2 (10^{-10}) \mathrm {\,MPa}$ \textrm {and} $m=16$}
\begin{figure}[h]
   	\centering
   	\includegraphics[scale=0.5]{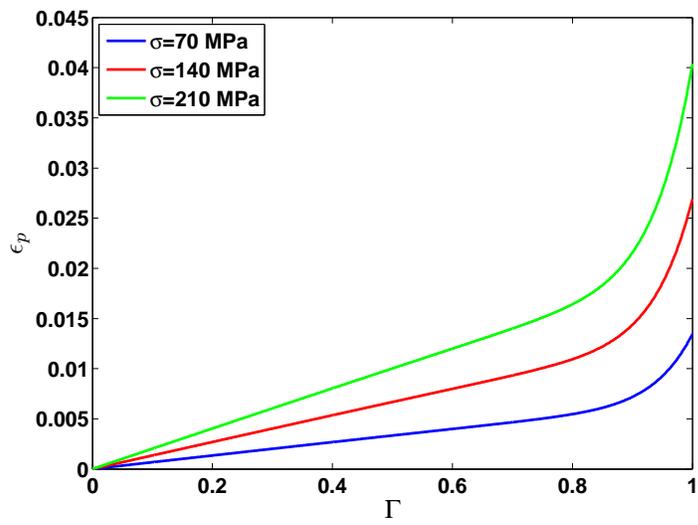}
   	\caption{Creep strain with $\Gamma$ for different $\sigma$ and $m=16$. \label{fig20}}
   \end{figure}

\begin{figure}[h]
   	\centering
   	\includegraphics[scale=0.5]{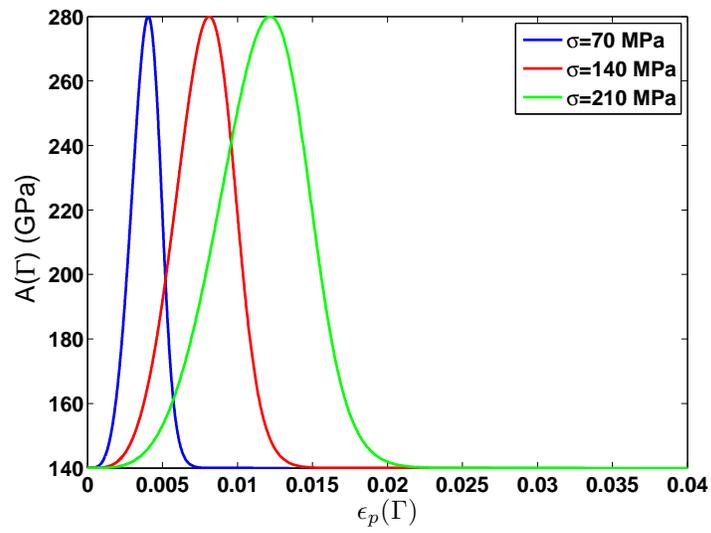}
   	\caption{Nonlinearity parameter $\mathrm A(\Gamma)$ with creep strain for different $\sigma$ and $m=16$. \label{fig21}}
   \end{figure}

Here, we assume that $\mathrm W_{nel}(\Gamma)$ depends on stress in a quadratic fashion. Figure \ref{fig20} shows the creep-strain with $\Gamma$ for different `$\sigma$'. As can be seen, the creep-strain increases with increasing $\sigma$. Also, Figure \ref{fig21} shows the nonlinearity parameter with creep-strain for different `$\sigma$'. Note that the above results are for the specific choices made for the constitutive responses of $\mathrm W_{nel} (\Gamma)$ and $\mathrm A(\Gamma)$. In a general context, one can choose the response functions for $\mathrm W_{nel} (\Gamma)$ and $\epsilon_p(\Gamma)$ and obtain $\mathrm A(\Gamma)$ or one can choose $\epsilon_p(\Gamma)$ and $\mathrm A(\Gamma)$ and obtain $\mathrm W_{nel} (\Gamma)$ from Eqns (\ref{eqn25a} \& \ref{eqn25b}). 

We would like to reiterate that the key advantage of the framework is that it results in damage evolution equations for macroscopic measurables like $\epsilon_p(\Gamma)$. However, the framework requires that one assume the relevant constitutive response functions, namely $\mathrm W_{el}(\bf E, \Gamma)$and $\mathrm W_{nel}(\Gamma)$. In many cases, the above choice is motivated by the experimental data as we demonstrated for the creep-like response where we chose the nonlinearity parameter ($\mathrm A(\Gamma)$) to be non-monotonic with $\Gamma$. Thus, this is an experimentally motivated material-state-based thermodynamic framework for damage prognosis.          
 
\section{Conclusion}
We presented a constitutive framework based on an internal variable approach to model the nonlinear elastic response of the materials undergoing progressive damage. The model is based on the construction of pseudo-elastic strain energy function that characterizes the thermodynamic response of the material undergoing degradation. The pseudo-elastic strain energy function is composed of $\mathrm W_{el}(\bf E, \Gamma)$ and $\mathrm W_{nel}(\Gamma)$. While $\mathrm W_{el}(\bf E, \Gamma)$ characterizes the elastic response of the material from a given material state $\Gamma$, $\mathrm W_{nel}(\Gamma)$ characterizes the energy dissipation through non-elastic processes in the material. A thorough discussion on the choice of constitutive forms for $\mathrm W_{el}(\bf E, \Gamma)$ and $\mathrm W_{nel}(\Gamma)$ was presented. Also, relevant experimental motivation behind their choice was highlighted. Specific cases included monotonic nonlinear elastic response for modeling fatigue behavior and non-monotonic nonlinear elastic response for modeling creep behavior in metals. Damage induced anisotropic nonlinear elastic response was qualitatively discussed. Then, we discussed a variety of constitutive responses for $\mathrm W_{nel}(\Gamma)$ along with their applicability under different damage progression scenarios. Finally, we presented two examples where progressive degradation was modeled under the thermodynamic framework presented in this article. These are:
\begin{enumerate}
	\item Stress-relaxation response in a spring-mass system
	\item Creep-like degradation
\end{enumerate}      
Stress-relaxation response of a spring-mass system was discussed for two cases --- one where the degradation is complete i.e., $\Gamma$ attains 1 and the other where the degradation is never fully completed but $\Gamma$  asymptotically approaches 1 i.e., $\Gamma \rightarrow 1$. For each $\Gamma$, the stiffness response of the spring was assumed to be bilinear and the second harmonic generation from small-amplitude vibrations of the mass about the equilibrium position was studied. It was found that as $\Gamma$ increases, the time-domain response of the displacement of the mass becomes asymmetric and also has even (second) harmonic frequency components. Moreover, the amplitude of the second harmonic was found to monotonically increase with $\Gamma$.   

Creep-like degradation response of material was studied. The nonlinearity parameter $\mathrm A(\Gamma)$ was assumed to be non-monotonic as observed in experiments. Governing equations of equilibrium were obtained from variational principles. These equations (ODE's) were solved to obtain accumulated creep (plastic) strain ($\epsilon_p(\Gamma)$). It was found that the creep strain monotonically increases with $\Gamma$ and more importantly, the nonlinearity parameter $\mathrm A(\Gamma)$ follows a non-monotonic trend with the creep strain i.e., $\epsilon_p(\Gamma)$.  

To summarize, the main features of the proposed framework are
\begin{enumerate}
	\item It is based on the notion of material state as identified by the internal variables.
	\item It is experimentally motivated and has a thermodynamic basis where the energetics of the damage progression are captured using the pseudo-elastic strain energy function.
	\item It results in a material-state-based damage prognosis approach particularly suitable for nonlinear ultrasonics.
\end{enumerate}  
Moreover, it should be possible to extend the framework to scenarios where multiple damage mechanism are in play by appropriately considering the energetics associated with each mechanism. However, it should be noted that this article restricts itself to cases where damage progression is assumed to be quasi-static. Our future work aims to address this aspect by considering explicit time-dependence of damage evolution. We also would like to extend the framework for degradation involving coupled processes like thermo-mechanical and chemo-mechanical degradations. 
       
 \section*{Acknowledgments}
The author would like to thank Prof. Cliff J Lissenden and Prof. Francesco Costanzo, Penn State, for their invaluable comments on an earlier version of the manuscript.
 
 \newpage	
   
   \clearpage
   
   \clearpage
   
%
%
%
%
%
%
%

%
%
%
%

\newpage

\bibliographystyle{elsarticle-harv}
\bibliography{Biblio-Database}
\end{document}